\documentclass[11pt]{article}
\usepackage{moriond,epsfig}

\def\be{\begin{equation}}
\def\ee{\end{equation}}
\def\bea{\begin{eqnarray}}
\def\eea{\end{eqnarray}}

\begin{document}
\vspace*{4cm}
\title{TOP QUARK MASS MEASUREMENTS}

\author{ L. CERRITO}

\address{Department of Physics, University of Illinois at Urbana-Champaign, 1110 West Green St.,\\
Urbana, IL 61801-3080 USA \\{\normalsize \rm on behalf of the CDF and D$\O$ Collaborations}}

\maketitle\abstracts{
Preliminary results on the measurement of the top quark mass at the Tevatron Collider are presented. In the dilepton decay channel, the CDF Collaboration measures $m_t$=175.0$^{+17.4}_{-16.9}$(stat.)$\pm$8.4(syst.) GeV/$c^2$, using a sample of $\sim$126 pb$^{-1}$ of proton-antiproton collision data at $\sqrt{s}=1.96$ TeV (Run II). In the lepton plus jets channel, the CDF Collaboration measures 177.5$^{+12.7}_{-9.4}$(stat.)$\pm$7.1(syst.) GeV/$c^2$, using a sample of $\sim$102 pb$^{-1}$ at $\sqrt{s}=1.96$$~$TeV. The D$\O$ Collaboration has newly applied a likelihood technique to improve the analysis of $\sim$125 pb$^{-1}$ of proton-antiproton collisions at $\sqrt{s}=1.8$ TeV (Run I), with the result: $m_t$=180.1$\pm$3.6(stat.)$\pm$3.9(syst.) GeV/$c^2$. The latter is combined with all the measurements based on the data collected in Run I to yield the most recent and comprehensive experimental determination of the top quark mass: $m_t=178.0\pm2.7({\rm stat.})\pm3.3({\rm syst.})$ GeV/$c^2$.}

\section{Introduction}
The discovery of the top quark at the Tevatron Collider \footnote{The Tevatron is a proton-antiproton synchrotron accelerator colliding beams at a center-of-mass energy of 1.96 TeV (Run II) in two locations (CDF and D$\O$). The Tevatron operated until 1998 (Run I) at a center-of-mass energy of 1.8 TeV.} in 1995 \cite{topdiscovery} marked the beginning of the successful measurement of many properties of the top. These include the production cross section \cite{topxsec} at $\sqrt{s}=1.8$ TeV, the helicity of the $W$ boson in top decays \cite{tophelic}, bounds on $t\bar{t}$ spin correlations \cite{tspincorr} and limits on the cross section of single top production \cite{singletop}. The value of the pole mass, ($m_t$), of the top quark was also measured to be \cite{toprun1}: 174.3$\pm$5.1 GeV/$c^2$. 

Such a large value of $m_t$, $\sim$35 times larger than the mass of the next heaviest quark, leads to several speculations about the nature of the top mass. Is $m_t$ really generated by the Higgs mechanism and is its value related to the top-Higgs-Yukawa coupling \cite{topmassspec} ? Moreover, a large $m_t$ gives the top particular relevance in the calculation of the parameters of the Standard Model (SM). In the SM electroweak theory, quantum corrections introduce a quadratic dependence to the $W$ boson mass on $m_t$ (and logarithmic on the mass of the Higgs boson, $M_H$), providing the opportunity for a stringent consistency test of the model. The uncertainty of such constraint on $M_H$ is today of $\sim$50\%, with uncertainties on the input $W$ boson mass less than 0.1\% and on the top quark mass of $\sim$3\%. 

\section{Top Quark Mass Measurements}
Top quarks are produced at the Tevatron in pairs of top-antitop ($t\bar{t}$) via strong interactions, with the cross section expected between 6.7 and 7.5 pb \cite{theoxsec1,theoxsec2} (NLO, for $m_t$=175 GeV and $\sqrt{s}~=~2.0$$~$TeV). 
The dominant production mechanism is quark-antiquark annihilation, while about 15\% is produced through gluon-gluon fusion. 
The decay of top quarks proceeds almost exclusively as $t\rightarrow Wb$, therefore it is natural to classify the final state of a $t\bar{t}$ event according to the decay modes of the $W$ boson. The decay channels are indicated as: {\it dilepton}, when both $W$'s decay leptonically, {\it lepton plus jets} when one of the $W$'s decays leptonically and the second decays hadronically, and the {\it all hadronic} channel, when both $W$'s decay hadronically. 
The dilepton decay channel yields the purest sample, with signal over background ($S/B$) on the order of $10:1$, but it accounts only for about 4/81 $\approx$ 4.9\% of $t\bar{t}$ decays. The lepton plus jets channel is the golden dataset for the measurement of the top mass at the Tevatron. It accounts for roughly 8/27 $\approx$ 30\% of the total decays, while mantaining $S/B\approx 1:1$. The all-hadronic decay channel accounts for about 44\% of the decays, but the sample is affected by a large QCD background, with $S/B$ on the order of $1:10$.

The presence of jets originated by a $b$ quark is often used to discriminate $t\bar{t}$ events against background. The identification of $b$ quarks in jets is made either through the measurement of a displaced secondary vertex in the event (due to the long lifetime of $b$-hadrons), or through the detection of a muon or electron from the semileptonic decay of $b$-hadrons.

\subsection{Dilepton Channel}
The CDF Collaboration has measured the mass of the top quark using the dilepton decay channel from a sample of  $\sim$126 pb$^{-1}$ $p\bar{p}$ collision data at $\sqrt{s}$=1.96 TeV. The sample was collected between May 2002 and July 2003 (Run II). Top candidate events are selected by requiring two isolated leptons (electrons or muons) with transverse momentum \footnote{The transverse momentum, $p_T$, is the projection of a particle's momentum on the plane perpendicular to the beam axis. The missing transverse energy, $E\!\!\!\!/ _T$, is identified with the undetected neutrino's transverse momenta. The pseudo-rapidity $\eta$ is defined from the polar angle $\theta$ (measured with respect to the proton direction) as $\eta=-{\rm ln}~{\rm tan}(\theta/2)$.} greater than 20 GeV/$c$, missing transverse energy greater than 25 GeV and at least two jets in the event. Jets are selected with energy greater than 10 GeV and pseudo-rapidity $|\eta|\leq$ 2.0. Since $t\bar{t}$ decays produce central spherical events with large total transverse energy ($H_T$), signal candidates are also required to have $H_T>200$ GeV. A total of 6 events are selected as $t\bar{t}$ candidates, while only 0.5$\pm$0.2 are expected from known sources of background. These include $WW$, Drell-Yan and $Z\rightarrow\tau\tau$ processes. The mass of the top is reconstructed on an event-by-event basis by solving the kinematic equations of a $t\bar{t}\rightarrow WbWb\rightarrow \ell\nu b \ell \nu b$ decay chain. Due to the presence of two neutrinos, this system is under constrained. One extra condition is imposed on the longitudinal momentum of the $t\bar{t}$ system, leaving only a twofold ambiguity in the neutrino momenta. For each event, the jet energies and $E\!\!\!\!/ _T$ are smeared according to the detector resolution, and the jets are associated to the $b$-quarks to reconstruct the decay chain. The twofold neutrino ambiguity is treated in the number of possible solutions. The configuration which yields a reconstructed top mass with a probability higher than the others is taken as the event configuration, and its reconstructed mass is chosen as the "raw mass" for the event. The final step is to fit the distribution of the raw mass from all events to a set of Monte Carlo-generated distributions with known input $m_t$. 
The best fit returns $m_t$=175.0$^{+17.4}_{-16.9}$(stat.)$\pm$8.4(syst.)~GeV/$c^2$. The systematic uncertainty includes a dominant contribution from the jet energy determination ($\approx$ 72\%), a contribution from $t\bar{t}$-event modelling ($\approx$ 25\%) and to a lesser extent it is affected by the uncertainty on the estimate and modelling of the background ($\approx$ 3\%). 

\subsection{Lepton Plus Jets Channel at CDF}
The CDF Collaboration has measured the mass of the top quark using the lepton plus jets decay channel from a sample of $\sim$102 pb$^{-1}$ $p\bar{p}$ collision data at $\sqrt{s}$=1.96 TeV. The sample was collected between March 2002 and May 2003 (Run II). Top candidate events are selected by requiring one, and only one, isolated electron or muon with $p_T>$20 GeV/$c$ and missing transverse energy in the reconstructed event greater than 20 GeV. At least four jets must be present, three of which with energy greater than 15 GeV, the fourth with energy greater than 8 GeV and all within $|\eta|<$2.0. In order to improve the signal purity, this analysis requires that at least one jet is identified as originating from a $b$-quark ($b$-jet) by detecting a displaced secondary vertex. A total of 22 events pass all the candidate selection requirements, with an expected 5.9$\pm$2.1 events coming from known background sources. These include a dominant contribution from $W$+jets processes (3.4$\pm$0.5 events), where one of the jets is mis-identified as a $b$-jet. Other contributions come from $Wb\bar{b}$, $Wc\bar{c}$, $Wc$ events. Minor sources of background are also due to $WW/WZ$, multijets and single top production.

The top mass reconstruction follows a set of steps similar to those described in the analysis of the dilepton channel. First, the kinematic equations of the $t\bar{t}$ decay chain are imposed. There are 24 configurations for each event, 12 from the assignment of the observed jets to the partons and 2 from a twofold ambiguity in the longitudinal component of the neutrino momentum. For each configuration, an event $\chi^2$ is calculated and minimized. The $\chi^2$ takes into account the detector resolution on the measured quantities as well as the $W$ boson and top-quark decay widths. The configuration that yields the lowest minimum $\chi^2$ is taken as the configuration for the event. The distribution of the reconstructed mass is compared and fit to Monte Carlo generated templates with known input $m_t$. The best fit returns 177.5$^{+12.7}_{-9.4}$(stat.)$\pm$7.1(syst.) GeV/$c^2$. 

The systematic uncertainty is dominated by the measurement of the jet energy (6.2 GeV/$c^2$). This in turn arises mainly from the calibration of the CDF calorimeter to a uniform response and setting the absolute scale of the jet energy measurement. The uncertainty associated with both these procedures will significantly be reduced in the future, also as a result of the increasing dataset of the samples used for the calibrations. The $t\bar{t}$ event modeling, which includes initial and final state radiation, parton distribution functions and MC generator, accounts for an uncertainty of about 3$~$GeV/$c^2$ on the measured $m_t$.

\subsection{Lepton Plus Jets Channel at D$\O$}
The D$\O$ Collaboration has successfully applied a likelihood technique to improve the statistical significance of an earlier measurement of the top mass in the lepton plus jets channel. The analysis uses $\sim$125 pb$^{-1}$ of data collected between 1994 and 1996. The previous measurement, based on a fit to the reconstructed top mass, determined \cite{d0run1}: $m_t$=173.3$\pm$5.6(stat.)$\pm$5.5(syst.)$~$GeV/$c^2$. The new approach \cite{likel} considers the probability density for each event:
\begin{equation}
P(x,m_t)=\frac{A(x)}{\sigma}\int d\sigma(y,m_t) dq_1 dq_2 f(q_1)f(q_2) W(y,x),
\end{equation}
where $d\sigma(y,m_t)$ is the differential (LO) cross section for $t\bar{t}$ production or for the background (the $W$+4 jets matrix elements are taken to model the background), the $f$ functions are the parton distributions and $W(y,x)$ represents the probability that the true value $x$ of an observable quantity is measured as $y$. The multiplicative factor $A(x)$ is the detector acceptance. The different jet-parton assignments and a twofold neutrino ambiguity are taken into account by summing $P(x,m_t)$ over the different possibilities. A log-likelihood number comprising the set of all candidate events is then maximized with respect to the top quark mass.

The selection of $t\bar{t}$ candidate events follows the same requirements of the previously published analysis with two differences. The first is that exactly four jets must be present in the event, in order to use the $W+4j$ matrix elements as a model for the background. This reduces the number of candidate events from the original 91 to 71. The second difference comes from the need to increase the signal purity. Monte Carlo studies have shown that this technique introduces a systematic bias to the measured $m_t$ of up to 2 GeV/$c^2$, when the background reaches 80\% of the total signal candidates. In order to minimise such effect, only events whose probability to originate from background is less than 10$^{-11}$ (according to the probability of being a $W$+jets as given by the VECBOS \cite{vecbos} differential cross section) are retained. With this additional requirements, there remain 22 $t\bar{t}$ candidate events while 10 events are expected from known background sources. The maximum likelihood for the set of selected events yields: $m_t$=180.1$\pm$3.6(stat.)$\pm$3.9(syst.)$~$GeV/$c^2$. The improved statistical significance over the previous measurement is equivalent to a twofold increase of the data sample.

\begin{figure}[t]
\begin{center}
\epsfig{file=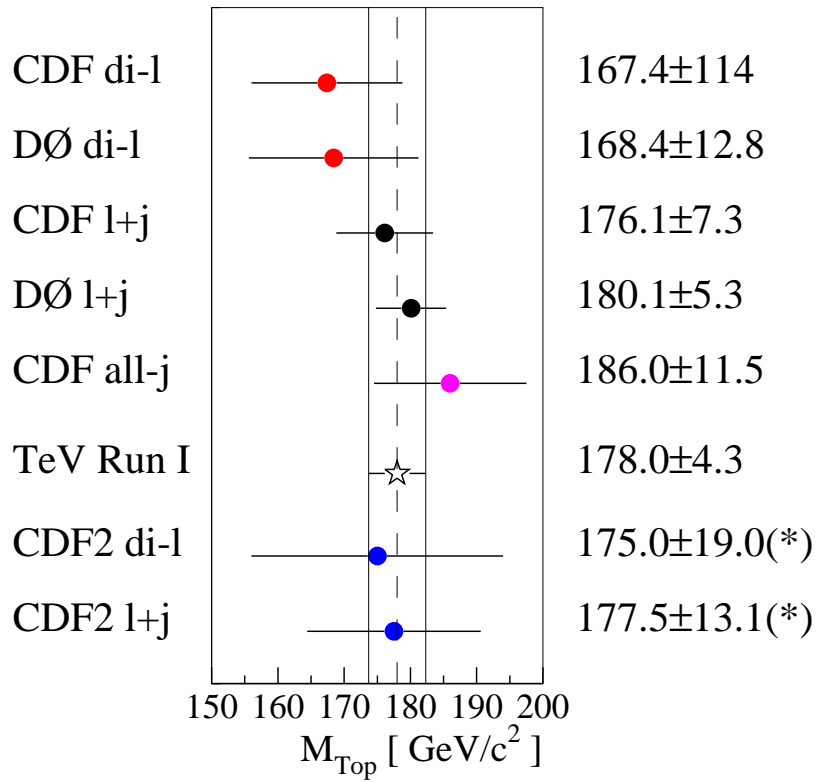,width=11cm,angle=0} 
\end{center}
\caption{Summary of the measurements of the top quark pole mass at the Tevatron. (*): Preliminary result.}
\label{fig:summary}
\end{figure}

\section{Combination of the Top-Quark Mass Measurements at the Tevatron}
The new measurement of the top pole mass in the lepton plus jets channel at D$\O$ warrants to revise the Tevatron Run I average $m_t$ previously calculated \cite{toprun1}. The combination includes measurements in the dilepton and lepton plus jets channel by CDF and D$\O$ and the measurement in the all hadronic channel by CDF (see Figure \ref{fig:summary}). Statistical and systematic uncertainties are properly taken into account. In particular, the statistical and fit errors are considered uncorrelated. The uncertainty due to the jet energy calibration is considered fully (100\%) correlated within measurements of each experiment. The modelling of the backgrounds is considered fully correlated within each decay channel and the signal modelling is considered fully correlated within all measurements. The new combined value for the top-quark mass is \cite{tevcombo2}: 
\begin{equation}
m_t=178.0\pm2.7({\rm stat.})\pm3.3({\rm syst.})~{\rm GeV}/c^2,
\end{equation}
or, combining statistical and systematic uncertainties: $m_t$=178.0$\pm$4.3 GeV/$c^2$. 

\section{Summary}
Preliminary results on the measurements of the top quark mass at the Tevatron are presented. Two new analyses in the dilepton and lepton plus jets decay channels, using 100$-$126$~$pb$^{-1}$ of data collected with the CDF detector,  establish a benchmark in the precision top mass measurement program for Run II at CDF. The D$\O$  Collaboration has applied a matrix elements likelihood technique to boost the statistical significance of the earlier measurement in the lepton plus jets channel using Run I data. When combined with other measurements of the two Tevatron experiments during the Run I period, this yields the most recent and comprehensive experimental measurement of the top quark pole mass: $m_t$=178.0$\pm$4.3 GeV/$c^2$. Many new results on the top mass determination are anticipated for the near future, together with a reduction of the systematic uncertainties, as different analysis techniques are refined and the Tevatron accelerator delivers an increasing dataset of proton-antiproton collisions. The Tevatron has delivered to date $\sim$500 pb$^{-1}$ of collision data and plans call for a dataset of 2 fb$^{-1}$ per-experiment by the year 2007.

\end{document}